# The Star-Planet Activity Research CubeSat (SPARCS): A Mission to Understand the Impact of Stars in Exoplanets

David R. Ardila
Jet Propulsion Laboratory
4800 Oak Grove Dr.; MS 306-392; Pasadena, CA 91109; 626-428-1355
David.ardila@jpl.nasa.gov

Evgenya Shkolnik, Paul Scowen,
School of Earth and Space Exploration, Arizona State University
781 E. Terrace Mall, Tempe, AZ, USA 85287-6004; 480-965-5081
shkolnik@asu.edu

April Jewell, Shouleh Nikzad
Jet Propulsion Laboratory
4800 Oak Grove Dr.; MS 306-392; Pasadena, CA 91109; 818-354-4321
April.D.Jewell@jpl.nasa.gov

Judd Bowman, Michael Fitzgerald, Daniel Jacobs, Constance Spittler,
School of Earth and Space Exploration, Arizona State University
781 E. Terrace Mall, Tempe, AZ, USA 85287-6004; 480-965-5081
Judd.Bowman@asu.edu

Travis Barman, Sarah Peacock
Lunar and Planetary Lab, University of Arizona
1415 N 6th Ave, Tucson, AZ 85705; 520-621-6963
barman.lpl.edu@gmail.com

Matthew Beasley
Southwest Research Inc.
1050 Walnut St #300, Boulder, CO 80302; 210-684-5111
beasley@boulder.swri.edu

Varoujan Gorgian
Jet Propulsion Laboratory
4800 Oak Grove Dr.; MS 306-392; Pasadena, CA 91109; 818-354-4321
varoujan.gorjian@jpl.nasa.gov

Joe Llama
Lowell Observatory
1400 W Mars Hill Rd, Flagstaff, AZ 86001; 928-774-3358
joe.llama@lowell.edu

Victoria Meadows
Dept. of Astronomy, University of Washington
3910 15th Ave NE, Seattle WA 98195-0002; 206-543-2888
vsm@astro.washington.edu

Mark Swain, Robert Zellem
Jet Propulsion Laboratory
4800 Oak Grove Dr.; MS 306-392; Pasadena, CA 91109; 818-354-4321
mark.r.swain@jpl.nasa.gov

**ABSTRACT**

The Star-Planet Activity Research CubeSat (SPARCS) is a NASA-funded astrophysics mission, devoted to the study of the ultraviolet (UV) time-domain behavior in low-mass stars. Given their abundance and size, low-mass stars are important targets in the search for habitable-zone, exoplanets. However, not enough is known about the stars flare and quiescent emission, which powers photochemical reactions on the atmospheres of possible planets. Over its initial 1-year mission, SPARCS will stare at ≈10 stars in order to measure short- (minutes) and long- (months) term variability simultaneously in the near-UV (NUV – $\lambda_c$ = 280 nm) and far-UV (FUV – $\lambda_c$ = 162 nm). The SPARCS payload consists of a 9-cm reflector telescope paired with two high-sensitivity 2D-doped CCDs. The detectors are kept passively cooled at 238K, in order to reduce dark-current contribution. The filters have been selected to provide strong rejection of longer wavelengths, where most of the starlight is emitted. The payload will be integrated within a 6U CubeSat to be placed on a Sun-synchronous terminator orbit, allowing for long observing stares for all targets. Launch is expected to occur not earlier than October 2021.

## CUBESATS FOR ASTROPHYSICS

Payloads performing astrophysics observations from space need to overcome unique challenges. Astrophysics targets tend to be faint, which makes high sensitivity and payload stability crucial. Because targets can be distributed all over the sky, parts and systems need to survive in space for at least 1 year. Most importantly, astrophysics places a premium in new discoveries, as opposed to providing continuous operational or tactical data, or preserving the data record. Within the NASA landscape, every new astrophysics mission needs to demonstrate that it can produce discoveries that cannot be realistically done any other way.

Time-domain observations provide a unique niche in which small, dedicated, and relatively inexpensive astrophysics payloads can produce competitive science.[1,2] Although all astrophysical phenomena change in time, astronomers know very little about the time-domain: most astrophysics observations from space tend to be single 'snapshots', and not 'movies'. The very successful Kepler/K2 missions, which have discovered thousands of planets and transients, are examples of the power of the time-domain. Particularly when performed at wavelengths that cannot be reached from the ground, time-domain observations can provide a unique understanding of the physical processes at play.

Here we present SPARCS, the Star-Planet Activity Research CubeSat. SPARCS will use passively cooled, high-sensitivity UV detectors, a well-characterized pointing system, and a specialized orbit to perform long-term UV monitoring of low-mass stars and understand the impact of this emission in possible extrasolar planets.

SPARCS was selected for funding as part of NASA's 2016 Astrophysics Research and Analysis (APRA) opportunity. The mission is led by Arizona State University (PI Evgenya Shkolnik). NASA's Jet Propulsion Laboratory provides the instrument camera. ASU is in charge of payload Integration and Testing (I&T) and Operations. The mission is currently in Phase B, working towards Preliminary Design Review (PDR), and a launch date not earlier than October 2021. The launch opportunity has not been identified.

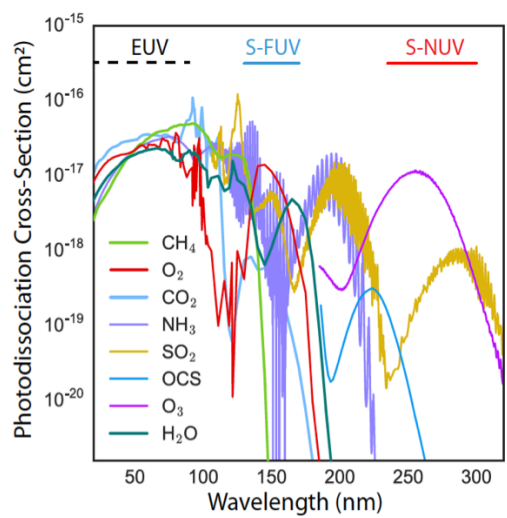

**Figure 1: The far and near UV (122 nm to 400 nm) contains photodissociation cross-sections peaks for important biosignatures. The SPARCS bands are indicated (S-FUV; S-NUV).**

## SPARCS SCIENCE

M-dwarfs are among the most common stars in the Galaxy. These are stars with masses ranging from 10% to 60% the mass of the Sun. Astronomers estimate that 40 billion of them may host at least one small planet in the "habitable zone"[3], the narrow zone around a star where liquid water could exist in the planet's surface. The small size of M-dwarfs, their large numbers, and the attendant abundance of planets around them, make these objects ideal targets for searches of Earth-like planets.

The UV radiation from M dwarfs is powerful and highly variable, persists at high levels for at least a ~Gyr and impacts planetary atmospheric loss, composition and habitability. The UV photons can dissociate molecules in the atmosphere of a possible planet (Figure 1), fundamentally altering its chemistry and the observables that we can use to determine habitability (Figure 2). The stellar UV radiation will impact the amount water, ozone, sulfur dioxide (a diagnostic of volcanic activity), and ammonia (an important source of the nitrogen required to build amino-acids) in the planetary atmosphere.

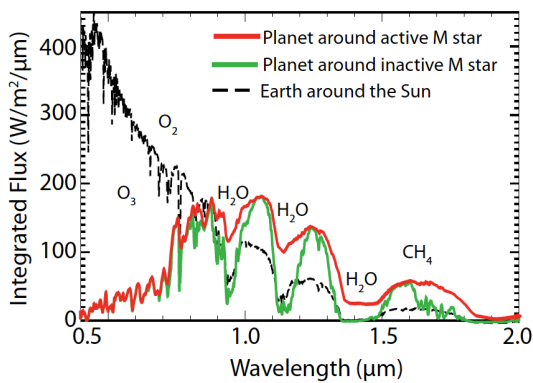

**Figure 2: The dramatic effect of stellar UV flux on an Earth-like planet's atmosphere orbing in the habitable zone of an active (high UV flux; red) and inactive (low UV flux; green) M4 dwarf. Adapted from [4].**

For these M-dwarfs the frequency and energy distribution of the UV flares, and their dependence with stellar mass and rotation period are not well-described. UV emission from stars does not reach the ground, and, while there have been a number of space-based, UV-capable missions (The International Ultraviolet Explorer - IUE, The Galaxy Evolution Explorer - GALEX, The Hubble Space Telescope - HST) none has devoted significantly large amounts of time to explore the time-domain UV emission of low-mass stars.

In this context, SPARCS' objectives are to measure the short- and long-term variability M-stars in the near-UV (NUV – 280 nm, centered on the MgII line) and far-UV (FUV – 162 nm, encompassing the CIV and HeII lines), to measure flare frequency and energy, as well as quiescent (non-flare) emission. SPARCS's two photometric bands have been selected to sample different emission regions within the stellar atmosphere (i.e. the chromosphere and transition region), and will provide crucial inputs to stellar models. These models in turn will be used to predict the observed spectra on extrasolar planets, and inform observations with facilities such as the James Webb Space Telescope.

We have designed a reference mission using a known set of targets. These encompass active and inactive, young and old, low mass stars of spectral types M0 to M6, ranging in brightness from 1 to 10 mJy at 280 nm. The target list will be complemented with objects discovered by the time of SPARCS launch, likely including results from the Transiting Exoplanet Survey Satellite (TESS). During its 1-year funded mission, SPARCS will observe ≈10 targets.

**SPARCS PAYLOAD**

The payload was designed to be compact, to minimize the number of reflections and transmissions, and to have no moving parts, in order to reduce cost and risk. It consists of a 9-cm telescope, a dichroic, two UV detectors, and associated electronics (See Figure 3).

The telescope is an f/4.4, two-mirror Ritchey-Chretien system. The mirror surfaces are protected with a layer of $MgF_2$ over the aluminum coating. The magnesium fluoride overcoat prevents the formation of the oxide layer and results in UV reflectivities larger than ≈80% at the SPARCS bands. The telescope system is constructed from low CTE materials in order to maintain focus during operations between -40 to 60 C.[5]

Light from the telescope is split by the dichroic optic at 233 nm. The transmitted beam is focused in the NUV channel, while the reflected beam is directed to the FUV channel. This allows simultaneous measurements in both bands.

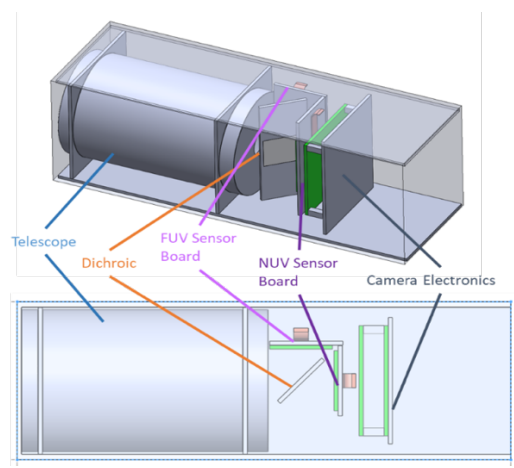

**Figure 3: Payload configuration.**

*Detectors and Filters*

SPARCS will serve to demonstrate the performance of high-sensitivity UV "2D-doped" (delta- or superlattice-doped) detectors in an operational environment[6].

The SPARCS camera (SPARCam) consists of two high quantum efficiency, UV-optimized, backside-illuminated detectors. Both are 2D-doped, AR-coated, 1k x 1k CCDs (Teledyne-e2v CCD47-20). The 2D-doping processes places a thin layer of atoms on the back-surface of the detector, preventing the formation of Si/SiO2 traps. The process results in the silicon detectors reaching the theoretical maximum internal QE[11]. The addition of an anti-reflection coating results in an overall quantum efficiency values (QE) larger than ≈30%. SPARCS will serve to demonstrate the performance of these detectors in an operational environment.

The impact of these high sensitivity detectors cannot be overstated. In the NUV channel, SPARCS has an effective area ≈20% that of GALEX, although the aperture area ratio between the two observatories is 3%. The 2D-doped detector technology is what makes SPARCS possible.

2D-doped detectors are in use in facilities such as the Wide field Prime Focus Camera for Palomar, and Zwicky Transient Facility, and were used in the Faint Intergalactic Red-shifted Emission Balloon (FIREBall-2) experiment. SPARCS will be the first long-duration space mission to use them.

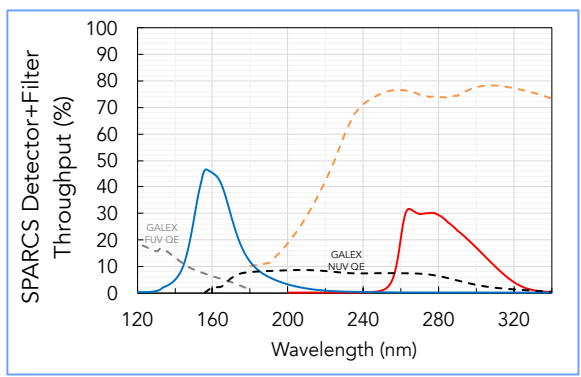

**Figure 4: Filter and detector responses. The baseline S-FUV (blue) and S-NUV (red) filter responses are indicated. The S-NUV response is the product of the 2D-doped detector response (dotted orange line) and the commercial filter response (not shown). The S-FUV filter is directly deposited on the detector's surface.**

*Red Rejection*

When not flaring, the target stars emit most of their light in the visible and infrared (IR) spectral regions. The emission at 1 μm is 500x (1200x) larger than the UV emission at 280 nm (162 nm) for the warmer low-mass stars. Any filter to be used in the UV needs to suppress long-wavelength contributions.

For the NUV we have baselined a commercial filter with FWHM≈40 nm, placed directly in front of the detector. Overall system transmission at 1 μm is 5 orders of magnitude smaller than in the NUV.

In the FUV, the lack of viable commercial alternatives led us to baseline a JPL-developed Metal Dielectric Filter (MDF) directly deposited on the detector[7]. The MDF consists of reflecting metal layers separated by a transparent spacer layer in order to destructively phase-match the reflection from the metal layer at a particular wavelength and maximize transmission through the structure[6]. For SPARCS, the 7-layer Al/AlF3 filter stacks optimized for the 160 nm bandpass provides in-band transmission of ≈30% at the peak, and 4 orders of magnitude of red-leak suppression.

*Imaging Characteristics and Jitter*

SPARCS will measure brightness changes in point sources on two UV bands. Absolute calibration will be done with well-known UV flux standards, such as white dwarfs. Other objects in the field will be used to perform relative calibration and to advance ancillary science programs. The size of the image on the detector impacts the spatial resolution and the sensitivity.

The size and shape of a point source in the detector will be dominated by jitter. These high-frequency pointing errors result in a blur spot analogous to the effect of ground-based atmospheric seeing. The exact impact of jitter depends on the type of Attitude Control System (ACS), the shape and mass of the spacecraft, and the location of the star tracker with respect to the payload optical path. A smaller jitter allows for better spatial resolution and higher sensitivity.

We expect SPARCS' jitter to be $1\sigma \approx 6"$ over 10 minutes. The ASTERIA mission has demonstrated $1\sigma \approx 0.5"$ pointing jitter using Blue Canyon Technology's (BCT) XACT unit in a closed feedback loop with a piezo-controlled focal plane[9]. However, this requires fast detector readouts (20 Hz, in the case of ASTERIA), which is only possible with bright targets. When using the ACS in coarse pointing mode (no payload feedback), ASTERIA has demonstrated jitter values $1\sigma \approx 5"$[10].

The blur spot will be Nyquist-sampled. While the nominal FOV is 2° in size, the images will be corrected for aberrations only to 1° diameter FOV.

**SPARCS CONCEPT OF OPERATIONS**

SPARCS will be launched to a Sun-synchronous terminator orbit on an initial 1-year mission. At the baseline altitude of 550 km, the spacecraft will only experience solar eclipses for ≈25 minutes per year. The terminator orbit allows for observations over long periods of time and is ideal for time-domain experiments.

The ecliptic latitude of the targets is constrained primarily by the spacecraft power system. For the baseline design, solar β-angles (between the solar array normal and the Sun) as large as 40° are possible. This allows coverage of 60% of the sky over a year.

The availability of each target during the year depends on its position in the sky and the time of the year the observation takes place. Targets in the ecliptic plane are available for periods of time of ≈3 months.

Each target will be observed over 1 to 3 stellar rotations. These range from 1 day to 80 days. The targets will be observed simultaneously with both bands, at exposure times large enough to measure their known variability changes[8]. Typical individual exposure times are ≈10 min.

As the spacecraft goes around the Earth, stellar observations are interrupted by target eclipses, and by the maximum β-angles allowed (Figure 5). Even with this, orbit allow for very large potential observing durations (Figure 5).

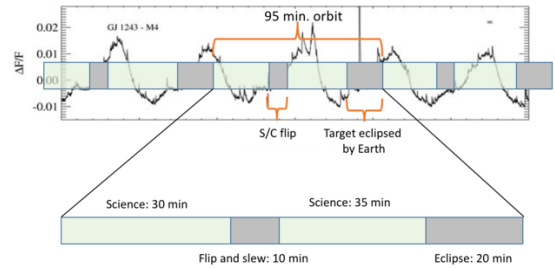

**Figure 6: A possible observation pattern over three spacecraft orbits. The black trace shows a notional stellar light curve, with sharp flares superimposed to the rotational variability. In this example, the star is available for an observation for 75 min, before being eclipsed by the Earth. The observing duration is further interrupted by the need to flip the spacecraft to point the radiator to space**

Stellar flaring may result in brightness changes as large as 1000x. To avoid saturation, SPARCS will use a dynamical exposure scheme in which the stellar brightness will be measured after each observation and the exposure time adjusted to avoid saturation.

The payload will extract postage stamps around each source of interest in the image field. These include the target star, relative calibrators, and objects associated with ancillary science programs.

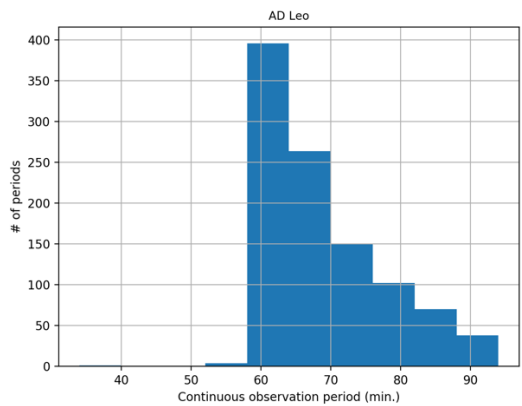

**Figure 5: Potential observing durations available for AD Leo throughout a year. The histogram includes constraints due to target eclipses and β-angle limits. In the case of AD Leo there is also a time during the year in which the star can be observed for ≈10 days without interruptions.**

As a practical matter, these observing durations will need to be interrupted to perform thermal management maneuvers and communication slews (Figure 6).

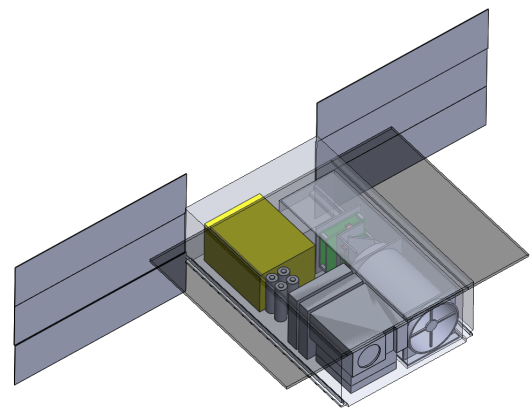

**Figure 7: Notional model for SPARCS' baseline bus architecture. The payload consists of a telescope and camera, which occupy a 3U volume. The star tracker is assumed to be co-boresighted with the payload telescope. Thermal control is achieved by a deployable radiator, as indicated.**

While some basic on-board data processing is necessary to measure the stellar brightness, most processing takes place on the ground. The ground science data system will produce calibrated light-curves for each target in each band. These can be then used to measure the flare frequency distribution and map the stellar brightness as a function of stellar rotation phase.

## SPACECRAFT BUS

The SPARCS bus baseline architecture assumes a 6U CubeSat in which the payload (telescope, dichroic, camera, associated electronics) is allocated 3U of space. The current architecture places the star tracker co-boresighted with the payload telescope (Figure 7). Notional subsystem interfaces are shown in Figure 8.

Power is provided by 0.2 m$^2$ of fixed, deployable solar arrays, and a bank of 6 Li-ion batteries. SPARCS total power consumption during science operations is <35W.

The communication system design is a UHF-uplink and S-Band downlink system, with an S-Band patch antenna and a UHF customized antenna. With an average of 2 passes a day, the system can support 190 MB/day. SPARCS data will be in the form of multiple postage stamps (≈10 pix x ≈10 pix), which reduces the required data volume to ≈120 MB/day.

The attitude control system requirements are based on the need to keep the high-frequency pointing errors at values that are consistent with the telescope optical prescription. The ACS system will use torque rods to de-saturate the reaction wheels.

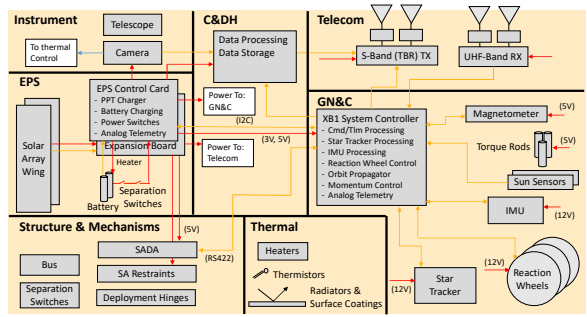

**Figure 8: SPARCS spacecraft block diagram. The functional subsystems for the mission design are shown with notional constituent elements and expected interfaces between those subsystems. In this diagram the Guidance Navigation and Control subsystem is baselined to have BCT's XB1 performance characteristics.**

In order to reduce the contribution of dark current noise to the observed photometry, the detector needs to be kept at 238K. This is made possible by a 1200 cm$^2$ radiator. The total area requires a deployable surface to be pointed to cold space during observations. This thermal solution requires that the spacecraft is rotated every half an orbit.

## OVERCOMING CHALLENGES

SPARCS' architecture is well-matched to the science goals and serves to showcase the power of CubeSats in advancing competitive astrophysics science. In particular, SPARCS is made possible by:

- High sensitivity UV detectors: The surface doping process allows standard CCDs to become sensitive UV detectors. Current CCDs are low-cost, low-power, high-reliability elements, and their use reduces the risk to the overall mission. The 2D-doping process allows for higher sensitivity with a smaller aperture.
- Solar-blind filters: SPARCS targets are very faint in the UV, compared to the visible and IR. Red-rejection ("Solar-blind") filters are crucial to control the noise within the band of interest. This is not an issue for naturally solar-blind UV detector technologies, such as microchannel plates, but is crucial to the use of 2D-doped detectors.
- Low jitter over long exposure times: SPARCS exposure times depend on the particular target but are typically ≈10 minutes. Over this period of time the high-frequency pointing errors need to be small, to reduce image blur and preserve sensitivity. A new generation of compact, affordable ACS units, based on star trackers, allows jitter values <10".
- Postage stamps to reduce the communication link capabilities: Because the science case is based on individual point sources and not on radiance measurements, the dataset is reduced by using postage stamps. This allows for standard communication links, again reducing risk and cost.
- Sun-synchronous terminator orbit: the orbit permits long stares per target, even allowing uninterrupted multi-day observations for fortuitously placed objects. Earth-trailing or L2 orbits would allow for even longer stares, but those seem to be beyond the reach of current CubeSats.

*Acknowledgments*

The SPARCS team acknowledges support from the NASA APRA program (NNH16ZDA001N-APRA). Research described in this paper was carried out in part at the Jet Propulsion Laboratory, California Institute of Technology, under a contract with the National Aeronautics and Space Administration.